\documentclass[twocolumn]{aastex61}
\received{---}
\revised{----}
\accepted{--- }          %\today}
\usepackage{color}
\usepackage{graphicx}
\usepackage{amssymb}
\usepackage{array}
\usepackage{color}
\usepackage{subfigure} 
\usepackage{natbib}
\bibpunct{(}{)}{;}{a}{,}{,}

\newcommand{\degree}{\ensuremath{^\circ}}

\def \sun {$_{\odot}$}

\shorttitle{Magnetic fields in Oph-B}
\shortauthors{Soam et al.}
\begin{document}

%\title{The JCMT BISTRO Survey: SCUBA-2 polarisation Measurements of Magnetic Fields in $\rho$ Oph-B region}

\title{Magnetic fields towards Ophiuchus-B derived from SCUBA-2 polarization measurements}

\correspondingauthor{Archana Soam}
\email{archanasoam.bhu@gmail.com}

\AuthorCallLimit=150

\author{Archana Soam}
\affiliation{Korea Astronomy and Space Science Institute, 776 Daedeokdae-ro, Yuseong-gu, Daejeon 34055, Republic of Korea}

\author{Kate Pattle}
\affiliation{Jeremiah Horrocks Institute, University of Central Lancashire, Preston PR1 2HE, UK}
\affiliation{National Astronomical Observatory of Japan, National Institutes of Natural Sciences, Osawa, Mitaka, Tokyo 181-8588, Japan}
\affiliation{Institute of Astronomy and Department of Physics, National Tsing Hua University, Hsinchu 30013, Taiwan}

\author{Derek Ward-Thompson}
\affiliation{Jeremiah Horrocks Institute, University of Central Lancashire, Preston PR1 2HE, UK}

\author{Chang Won Lee} 
\affiliation{Korea Astronomy and Space Science Institute, 776 Daedeokdae-ro, Yuseong-gu, Daejeon 34055, Republic of Korea}
\affiliation{Korea University of Science and Technology, 217 Gajang-ro, Yuseong-gu, Daejeon 34113, Korea}

\author{Sarah Sadavoy} 
\affiliation{Harvard-Smithsonian Center for Astrophysics, 60 Garden Street, Cambridge, MA 02138, USA}

\author{Patrick M. Koch}
\affiliation{Academia Sinica Institute of Astronomy and Astrophysics, P.O. Box 23-141, Taipei 10617, Taiwan}

\author{Gwanjeong Kim}
\affiliation{Korea Astronomy and Space Science Institute, 776 Daedeokdae-ro, Yuseong-gu, Daejeon 34055, Republic of Korea}
\affiliation{Korea University of Science and Technology, 217 Gajang-ro, Yuseong-gu, Daejeon 34113, Korea}
\affiliation{Nobeyama Radio Observatory (NRO), National Astronomical Observatory of Japan (NAOJ), Japan}

\author{Jungmi Kwon}
\affiliation{Institute of Space and Astronautical Science, Japan Aerospace Exploration Agency, 3-1-1 Yoshinodai, Chuo-ku, Sagamihara, Kanagawa 252-5210, Japan}

\author{Woojin Kwon}
\affiliation{Korea Astronomy and Space Science Institute, 776 Daedeokdae-ro, Yuseong-gu, Daejeon 34055, Republic of Korea}
\affiliation{Korea University of Science and Technology, 217 Gajang-ro, Yuseong-gu, Daejeon 34113, Korea}

\author{Doris Arzoumanian} 
\affiliation{Department of Physics, Graduate School of Science, Nagoya University, Furo-cho, Chikusa-ku, Nagoya 464-8602, Japan}

% ...

%\author{August Muench}
%\affiliation{American Astronomical Society \\
%2000 Florida Ave., NW, Suite 300 \\
%Washington, DC 20009-1231, USA}
%\collaboration{(AAS Journals Data Scientists collaboration)}

%\author{Butler Burton}
%\affiliation{National Radio Astronomy Observatory}
%\affiliation{AAS Journals Associate Editor-in-Chief}
%\nocollaboration

%\altaffiliation{Creator of AASTeX v6.1}
%\collaboration{(LaTeX collaboration)}

%\affiliation{AAS Director of Publishing}
%\affiliation{American Astronomical Society \\
%2000 Florida Ave., NW, Suite 300 \\
%Washington, DC 20009-1231, USA}




\begin{abstract}
We present the results of dust emission polarization measurements of Ophiuchus-B (Oph-B) carried out using the Submillimetre Common-User Bolometer Array 2 (SCUBA-2) camera with its associated polarimeter (POL-2) on the James Clerk Maxwell Telescope (JCMT) in Hawaii. This work is part of the B-fields In Star-forming Region Observations (BISTRO) survey initiated to understand the role of magnetic fields in star formation for nearby star-forming molecular clouds.  We present a first look at the geometry and strength of magnetic fields in Oph-B. The field geometry is traced over $\sim$0.2 pc, with clear detection of both of the sub-clumps of Oph-B. The field pattern appears significantly disordered in sub-clump Oph-B1. The field geometry in Oph-B2 is more ordered, with a tendency to be along the major axis of the clump, parallel to the filamentary structure within which it lies. The degree of polarization decreases systematically towards the dense core material in the two sub-clumps. The field lines in the lower density material along the periphery are smoothly joined to the large scale magnetic fields probed by NIR polarization observations.  We estimated a magnetic field strength of 630$\pm$410 $\mu$G in the Oph-B2 sub-clump using a Davis-Chandeasekhar-Fermi analysis. With this magnetic field strength, we find a mass-to-flux ratio $\lambda$= 1.6$\pm$1.1, which suggests that the Oph-B2 clump is slightly magnetically supercritical.

%. The observed chaotic magnetic fields in the region can be attributed to a distortion in the B-fields or modification in field lines by turbulent environment in the cloud region. 

\end{abstract}




\section{Introduction} \label{sec:intro}

Low mass stars are formed in dense cores ($\rm M\approx 1-10~M_{\odot}$, size$\approx$0.1-0.4 pc and density $\rm \approx 10^{4}-10^{5} cm^{-3}$) embedded in molecular clouds which are generally self-gravitating, turbulent, magnetized and thought to be compressible fluids and are expected to form one or a few stars when they become unstable to gravitational collapse.

Considering the magnetized nature of molecular clouds \citep{1987ARA&A..25...23S, 2007ARA&A..45..565M}, we expect magnetic fields (B-fields) to also have a significant impact on dense cores.  Nevertheless, the role of B-fields on the formation of dense cores and their evolution into the various stages of star formation is still under debate. Several observational and theoretical studies have been dedicated to understand the importance of B-fields in star formation. For example, isolated low-mass cores that are magnetically dominated may gradually condense out of a large scale cloud through ambipolar diffusion \citep[e.g.,][]{1987ARA&A..25...23S, 1993prpl.conf..327M, 1999osps.conf..305M, 2003ApJ...599..363A}. In this picture, the core will be flattened into a disk-like morphology on scales of a few thousand AU, with field lines primarily parallel to the symmetry axis.  These field lines become pinched into an hour-glass morphology as mass accumulates in the core and self-gravity becomes more significant \citep{1993ApJ...415..680F, 1993ApJ...417..243G, 2006Sci...313..812G, 2009ApJ...702.1584A}. On the other hand, B-fields are expected to be less significant if cores form via turbulent flows \citep{2004RvMP...76..125M, 2007ApJ...661..262D, 2010ApJ...723..425D}. In this picture the B-field morphology will be more chaotic \citep{2004Ap&SS.292..225C, 2017ApJ...842L...9H}. 

B-fields are often characterized by dust polarization observations. Dust grains are expected to align with their short axes parallel to an external B-field.  As a result, thermal dust emission at sub-millimeter or millimeter wavelengths from these grains will be polarized perpendicular to the B-field \citep[e.g.][]{1982MNRAS.200.1169C, 1984ApJ...284L..51H, 1988QJRAS..29..327H,1998ApJ...502L..75R, 2000ASPC..215...69L, 2000ApJS..128..335D, 2010ApJS..186..406D, 2012ApJS..201...13V, 2014ApJS..213...13H}. The actual mechanism by which dust grains align with a B-field is still unclear \citep{2007JQSRT.106..225L, 2015ARA&A..53..501A}. The most accepted mechanism to date is radiative torque alignment \citep[e.g.][]{2014MNRAS.438..680H, 2015MNRAS.448.1178H, 2015ARA&A..53..501A}, which was originally proposed by \citet{1976Ap&SS..43..291D}.

Polarized thermal dust emission is important because it probes the magnetic fields in the denser regions with extinction $\rm A_{V} >50$ mag, since NIR/optical polarization measurements are limited by the number of detectable background stars and cannot probe these high density environments. At sub-mm wavelength one can probe the B-field morphology deep inside the dense cores ($\mathrm{n_{H_{2}}\sim10^{5}-10^{6}cm^{-3}}$) where the central protostars and their circumstellar disks form. B-fields in more diffuse medium ($A_{V}\sim$ 1-20 mag), associated with larger-scale structures ($>$1 pc), can typically be probed using dust extinction polarization of background stars at optical or NIR wavelengths \citep{1949Sci...109..166H, 1949ApJ...109..471H, 1949Sci...109..165H}.

Ophiuchus is a molecular cloud at a distance of $\approx$140 pc \citep[e.g.][]{1981A&A....99..346C, 1989A&A...216...44D, 1998A&A...338..897K, 2004AJ....127.1029R, 2008ApJ...675L..29L, 2017ApJ...834..141O}. Therefore this is one of the closest low mass star forming regions \citep{1992lmsf.book..159W, 2000ApJ...545..327J, 2008ApJ...672.1013P, 2010A&A...518L.104M}. The Ophiuchus region is highly structured, with several $\sim$1 pc sized clumps. This study focuses on the Oph-B clump, which has two active star forming sub-regions named Oph-B1 and Oph-B2, separated by $\sim$5$\arcmin$. The Oph-B2 clump is the second densest region in Ophiuchus.

We present the first submm polarization observations made using the Submillimetre Common-User Bolometer Array 2 (SCUBA-2) camera with the POL-2 polarimeter towards the Oph-B region to map the B-fields on core scales. The molecular line observations towards this region using the Heterodyne Array Receiver Program (HARP)  are already available in \citet{2015MNRAS.447.1996W} to understand the kinematics of the cloud. 

The paper is organized as follows: in Section 2, we describe the observations and data reduction;  in Section 3, we give initial results; in Section 4, we analyze the polarization data and estimate the magnetic field strength; and in Section 5 we summarize the paper.

\begin{figure*}
\centering
\resizebox{10.5cm}{10cm}{\includegraphics{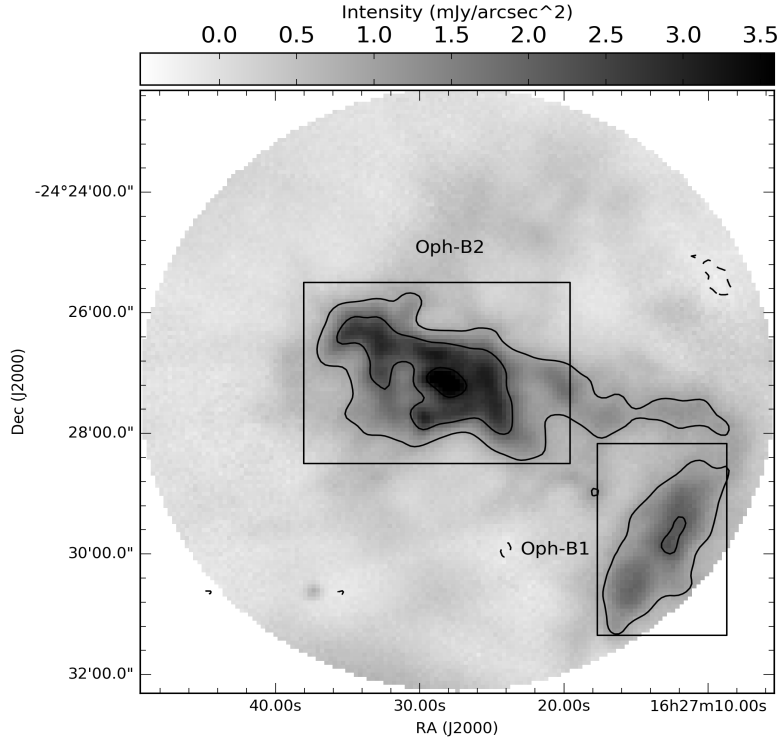}}
\caption{The 850$\mu$m dust emission map of Oph-B clump obtained using SCUBA-2 on the JCMT. The two sub-clumps, B1 and B2, are labeled with boxes. Contours are of 850$\mu$m emission (greyscale) and the contour levels are drawn with (0.6, 1.8, 2.9, 4.1) mJy/$\rm arcsec^{2}$.}\label{Fig:OphB_Imap}
\end{figure*}

\section{Data acquisition and reduction techniques} \label{sec:obs}

We observed Oph-B in 850-$\mu$m polarized emission with SCUBA-2 \citep{2013MNRAS.430.2513H} in conjunction with POL-2 \citep{2016SPIE.9914E..03F, 2018inpreparation} as part of the B-fields In STar-forming Region Observations \citep{2017ApJ...842...66W} survey under project code M16AL004 at the James Clerk Maxwell Telescope (JCMT).  This survey aims to use dust polarization maps of nearby molecular clouds to probe B-field structures.  Oph-B was observed over a series of 21 observations, with an average integration time of $\sim0.55$ hours per observation, in good weather (0.05 $<\tau_{225}<$ 0.08, where $\tau_{225}$ is atmospheric opacity at 225\,GHz).  Table \ref{tab:obslog} summarizes the observation logs. The scan pattern used was a POL-2 daisy \citep{2016SPIE.9914E..03F} producing a uniform, high signal-to-noise, coverage over the central 3${\arcmin}$ of the map.. This observing mode is based on the SCUBA-2 CV daisy scan pattern \citep{2013MNRAS.430.2513H} but modified to have a slower scan speed (8${\arcsec}$/s compared to 155${\arcsec}$/s) to obtain sufficient on-sky data for good Stokes Q and U values. This pattern maps a fully sampled, 12${\arcmin}$-diameter, circular region with an effective resolution of 14.1${\arcsec}$. Coverage decreases, with a consequent significant increase in the RMS noise, towards the edges of the map. The wave plate rotates at a frequency of 2 Hz. 

%==============
%\input{table1_log}		%Observation Log
\begin{table}
\begin{small}
  \caption{Log of SCUBA-2/POL-2 observations.}\label{tab:obslog}
  \begin{tabular}{lll}
\hline\hline
No. of Observations          &  Observation date  \\
                            & (year, month,date)                       \\
\hline
3              & 2016 Apr. 27        \\  
3              & 2016 Apr. 28           \\ 
5              & 2016 May 02         \\ 
1              & 2016 May 05           \\ 
5              & 2016 May 07          \\ 
2              & 2016 May 10           \\ 
2              & 2016 May 11            \\ 
\hline
\end{tabular}
\end{small}
\end{table}
%==============

The data were reduced using the {\tt\string pol2map} python script in the Starlink \citep{2014ASPC..485..391C} {\tt\string SMURF} \citep{2013ascl.soft10007J} package. The reduction followed three main steps. In step one, {\tt\string pol2map} uses the {\tt\string calcqu} command to create Stokes $Q$, $U$ and $I$ timestreams from the raw data.  It then creates an initial Stokes I map for each of the observations using the {\tt makemap} \citep{2013MNRAS.430.2545C} routine, and coadds these to create an initial estimate of the Stokes I emission in the region. In step two, {\tt\string pol2map} is re-run. The initial I map of the region is used to generate a fixed signal-to-noise-based mask which is used to define regions of astrophysical emission and thus to consistently re-reduce the previously-created Stokes I timestreams for each observation (see \citealt{2013MNRAS.430.2545C} for a detailed discussion on {\tt\string makemap} and \citet{2015MNRAS.454.2557M} for the role of masking in SCUBA-2 data reduction).  The new set of Stokes I maps are then coadded to produce the final Stokes I emission map of the region. In step 3, {\tt\string pol2map} is again re-run, this time creating a Stokes Q and U maps for each observation from their timestreams, using the same mask as used in step 2, and coadding these to produce final Stokes Q and U emission maps for the region.  The final Stokes Q, U and I maps, and their corresponding variance maps, are then used to create a polarization vector catalog. The vectors are de-biased to remove the effect of statistical biasing in low signal-to-noise-ratio (SNR) regions. See \citet{2015MNRAS.454.2557M} and \citet{2017ApJ...846..122P} for a detailed description of the SCUBA-2 and POL-2 data reduction process, respectively.

The RMS noise in the maps was estimated by selecting an emission-free region near the map center and finding the standard deviation of the measured flux density distribution in that region. A RMS noise value of $\sim$2 mJy/beam \citep{2017ApJ...846..122P} was set as the target value for the BISTRO survey. The estimated noise in the Stokes I map of Oph-B is $\rm \sim3.5~mJy/beam$. 

Bias in the measured polarization fraction and polarized intensity values result from the fact that polarized intensity is defined as being positive, and so uncertainties on Stokes Q and U values tend to increase the measured polarization values \citep{2006PASP..118.1340V, 2018arXiv180409313K}.  De-biased values of polarized intensity are estimated to be

\begin{equation}
\mathrm{PI = \sqrt{Q^{2}+U^{2}-0.5(\delta Q^{2}+\delta U^{2})}} ,
\end{equation}

where PI is polarized intensity, $\delta$Q is the uncertainty on Stokes Q, and $\delta$U is the uncertainty on Stokes U.  The polarization fraction, P, is then given by

\begin{equation}
\mathrm{P = \frac{PI}{I}}  ,
\end{equation}

where I is the total intensity.

%\begin{equation}
%\mathrm{P = \frac{\sqrt{Q^{2}+U^{2}}}{I}}\times 100%
%\end{equation}

The polarization position angle $\rm \theta$ and the corresponding uncertainties \citep{1993A&A...274..968N} are estimated as:

\begin{equation}
\mathrm{\theta = \frac{1}{2}tan^{-1}(\frac{U}{Q})}   ,
\end{equation}

and

\begin{equation}
\mathrm{\delta \theta= 0.5\times \frac{\sqrt{(Q^{2}\times \delta U^{2} +U^{2}\times \delta Q^{2})}}{(Q^{2}+U^{2})} \times \frac{180{\degree}}{\pi}}
\end{equation}

The debiased polarized intensity, polarization fraction and angle derived as explained above are used in the analysis of the paper. Magnetic field orientation is derived by rotating the $\theta$ by 90${\degree}$. The details of the procedure followed in producing polarization values is explained in detail by \citet{2018arXiv180409313K}.

\section{Results} \label{sec:results}

\begin{figure*}
\centering
%\resizebox{14cm}{12cm}{\includegraphics{spire500_scuba2_ovpl_YSOs.eps}}
\resizebox{14cm}{12cm}{\includegraphics{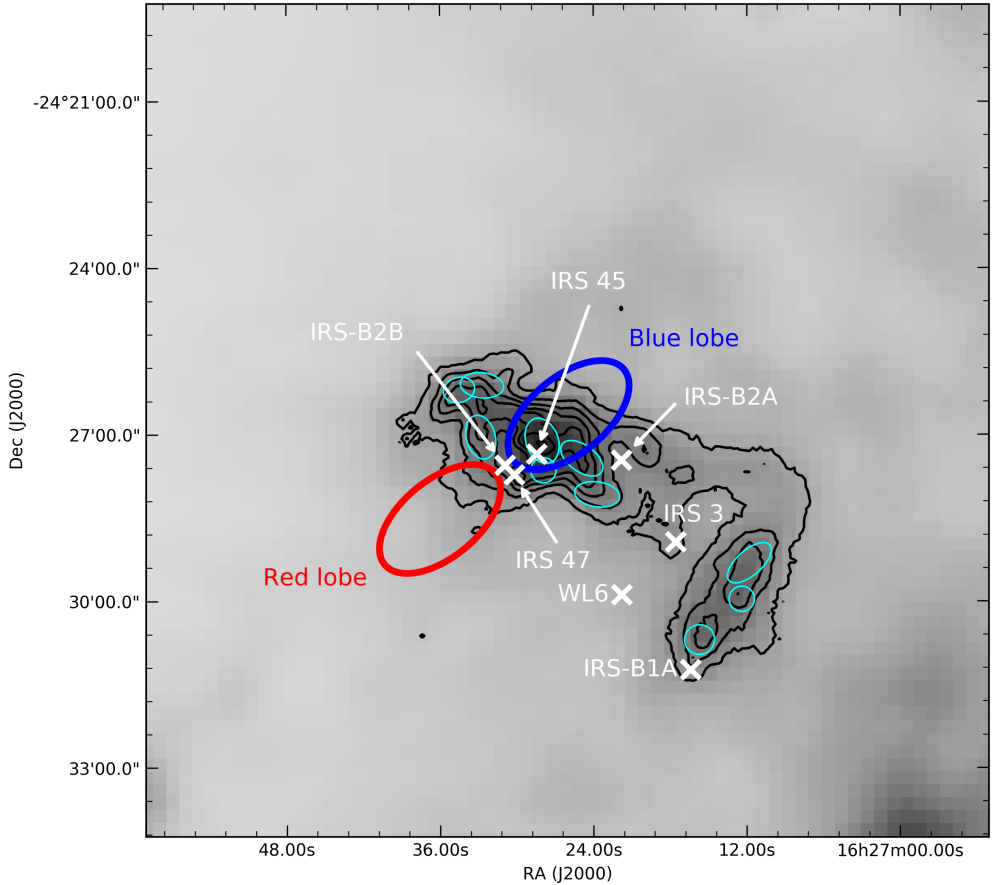}}
\caption{This figure shows submm continuum emission and outflow-related structures and YSO locations in Oph-B. The background image shows Herschel 500$\mu$m emission. Contours show the 850$\mu$m SCUBA-2 dust continuum emission. The positions of YSOs identified in the region are marked with white crosses and labeled. Blue and red ellipses show the blue- and red-shifted wings of the prominent outflow associated with IRS47 \citep{2015MNRAS.447.1996W}, respectively. Cyan ellipses show the dense structures identified in $\mathrm{N_2H^+}$ data \citep{2007A&A...472..519A}. IRS-B1A, IRS-B2A and IRS-B2B correspond to YSOs SSTc2dJ162716.4-243114, SSTc2d162721.8-242728 and SSTc2d162730.9-242733 in the SIMBAD database, respectively.}\label{Fig:YSOs}
\end{figure*}

Figure \ref{Fig:OphB_Imap} shows the Stokes I map of Oph-B obtained from SCUBA-2/POL-2 observations. The Oph-B clump contains two structures, one in the southwest known as Oph-B1 and other in the northeast known as Oph-B2 \citep{1998A&A...336..150M}. These structures are labeled with boxes in figure \ref{Fig:OphB_Imap}.  The entire clump appears to be rather cold, with dust temperatures ranging from $\sim$ 7 to 23 K at the center of the clump \citep{2007MNRAS.379.1390S}.

The Oph-B region contains at least three Class I protostars and five flat spectrum objects, many of which are driving outflows \citep{2015MNRAS.447.1996W}. Figure \ref{Fig:YSOs} shows the distributions of young stellar objects (YSOs) in Oph-B with the blue- and red-shifted lobes of the molecular outflows driven by IRS 47, as identified by \citet{2015MNRAS.447.1996W}, marked. Figure \ref{Fig:YSOs} also shows dense condensations identified using Gaussclump \citep{1990ApJ...356..513S} in $\mathrm{N_2H^+}$ (1-0) data by \citet{2007A&A...472..519A}.

%%IRS 54 is a wide binary system, separated by 7${\arcsec}$ based on near-IR observations \citep{2004A&A...427..651D, 2006AJ....132.2675H}, and as such, its outflow may be precessing \citep{2009A&A...507..861J}. 

\begin{figure*}%[t]
\includegraphics[width=90mm]{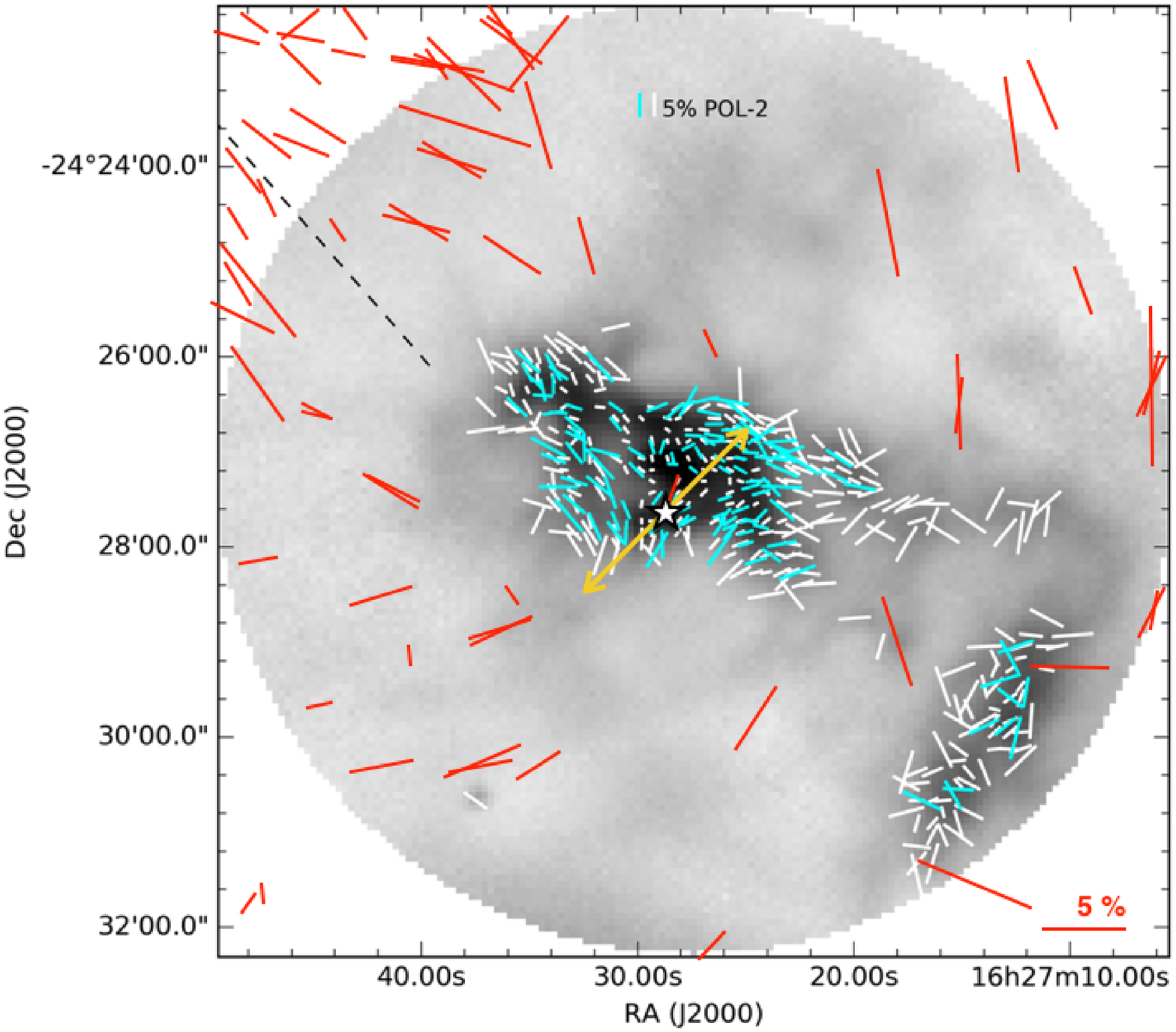}%
\includegraphics[width=90mm]{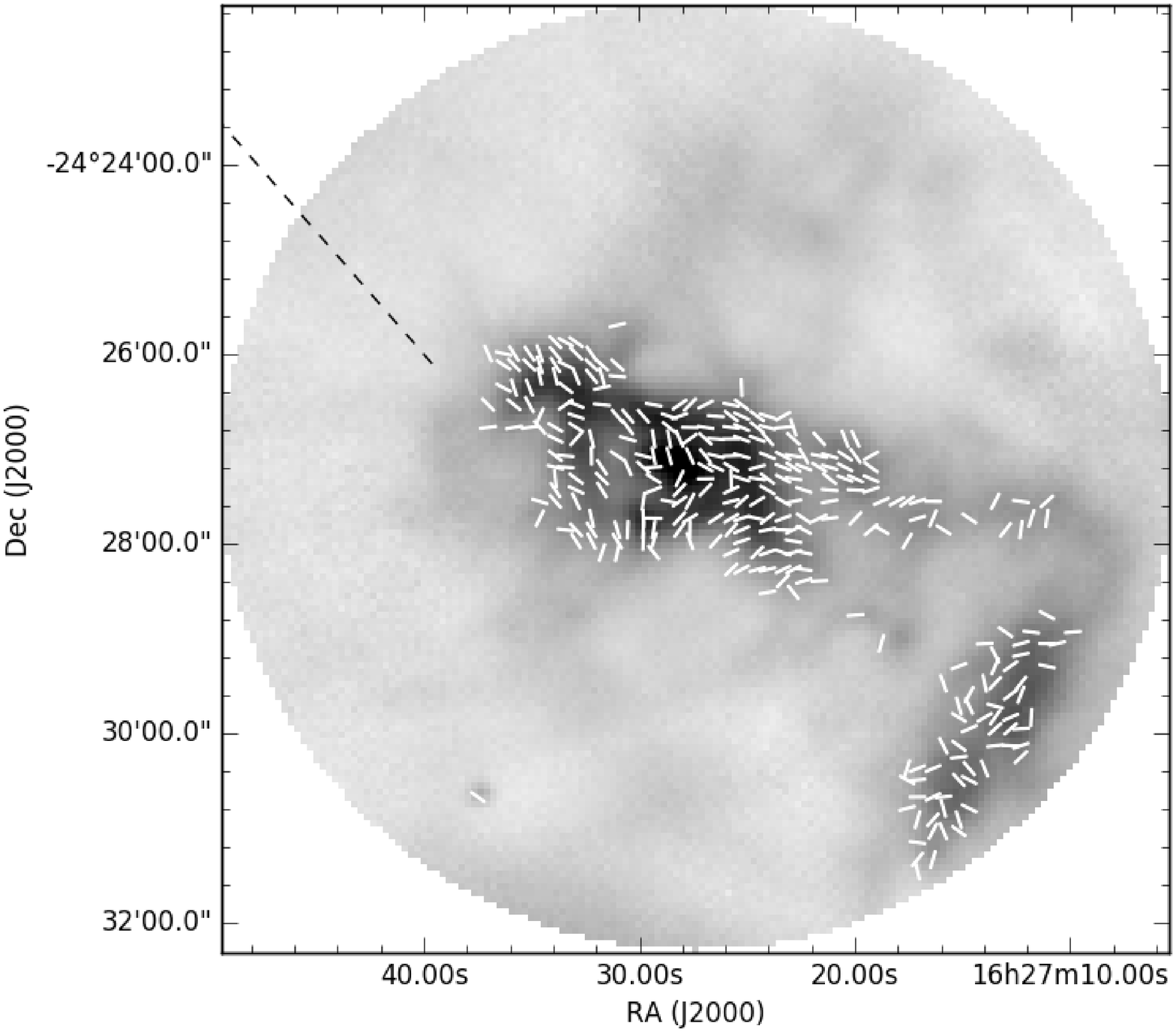}%
\caption{\textit{Left panel:} B-field orientation in Oph-B from 850$\mu$m dust polarization data. The background image shows the dust continuum map at 850$\mu$m from SCUBA-2. The white and cyan color vectors correspond to data with PI/$\delta PI >$ 2 and PI/$\delta PI >$ 3, respectively. The vectors are  rotated by 90$^{\degree}$ to show the inferred B-field morphology. The white star symbol represents the position of IRS47, with its associated bipolar outflow shown using a double-headed arrow in yellow. A 5\% polarization vector is shown for reference. The red vectors shows the B-fields mapped using deep NIR observations by \citep{2015ApJS..220...17K}. The black dashed line shows the orientation of the Galactic Plane at the latitude of the cloud. \textit{Right panel:} Same as left panel (except NIR data), but the vectors are of uniform length, rather than scaled with polarization fraction.}\label{Fig:pol_vec}
\end{figure*}

B-fields derived from dust polarization measurements are shown by white line segments in the left-hand panel of figure \ref{Fig:pol_vec}. The polarization vectors are rotated by 90$^{\degree}$ to show the B-field orientation. The length of the vectors shows the polarization fraction. The underlying image is the Stokes I map, showing the geometry of the cloud. For the following analysis we used those polarization measurements where $\frac{PI}{\delta PI}>$2 in order to ensure we have a large and robust sample. The measurements with $\frac{PI}{\delta PI}>$3 are also shown as cyan color vectors in figure \ref{Fig:pol_vec}. PI and $\delta {PI}$ are the polarized intensity and uncertainty on polarized intensity, respectively. The B-field orientations obtained from NIR polarization measurements by \citet{2015ApJS..220...17K} are shown in the figure as red line segments. The right-hand panel of figure \ref{Fig:pol_vec} shows the B-field orientation, with polarization vectors normalized. Here the lengths of vectors are independent of the polarization fraction.

\begin{figure}
\resizebox{8.5cm}{8cm}{\includegraphics{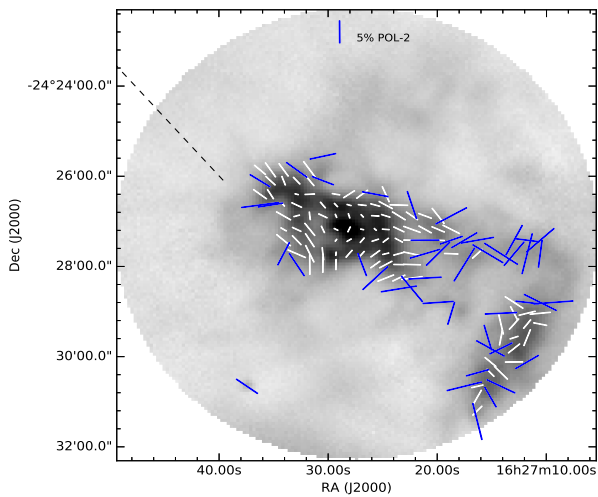}}
\caption{Smoothed B-field orientations in Oph-B, shown on the 850$\mu$m dust emission map from SCUBA-2. The polarization data are averaged over 4 pixels (an effective pixel size of 8$^{\prime\prime}$ using 2$\times$2 binning). White vectors correspond to data with P$<$5\%, while blue vectors correspond to data with P$>$5\%.  The vectors are selected to have PI/$\delta$PI $>$ 2, and are rotated by 90$^{\degree}$ to show the inferred B-field morphology. A reference vector with 5\% polarization is given for scale. The black dashed line shows the orientation of the Galactic Plane at the latitude of the cloud.}\label{Fig:binned}
\end{figure}

Figure \ref{Fig:binned} shows the polarization vectors (rotated by 90$^{\degree}$) with the data averaged over four pixels using 2$\times$2 binning, showing the general morphology of the B-fields. 

\begin{figure}
\resizebox{9cm}{9cm}{\includegraphics{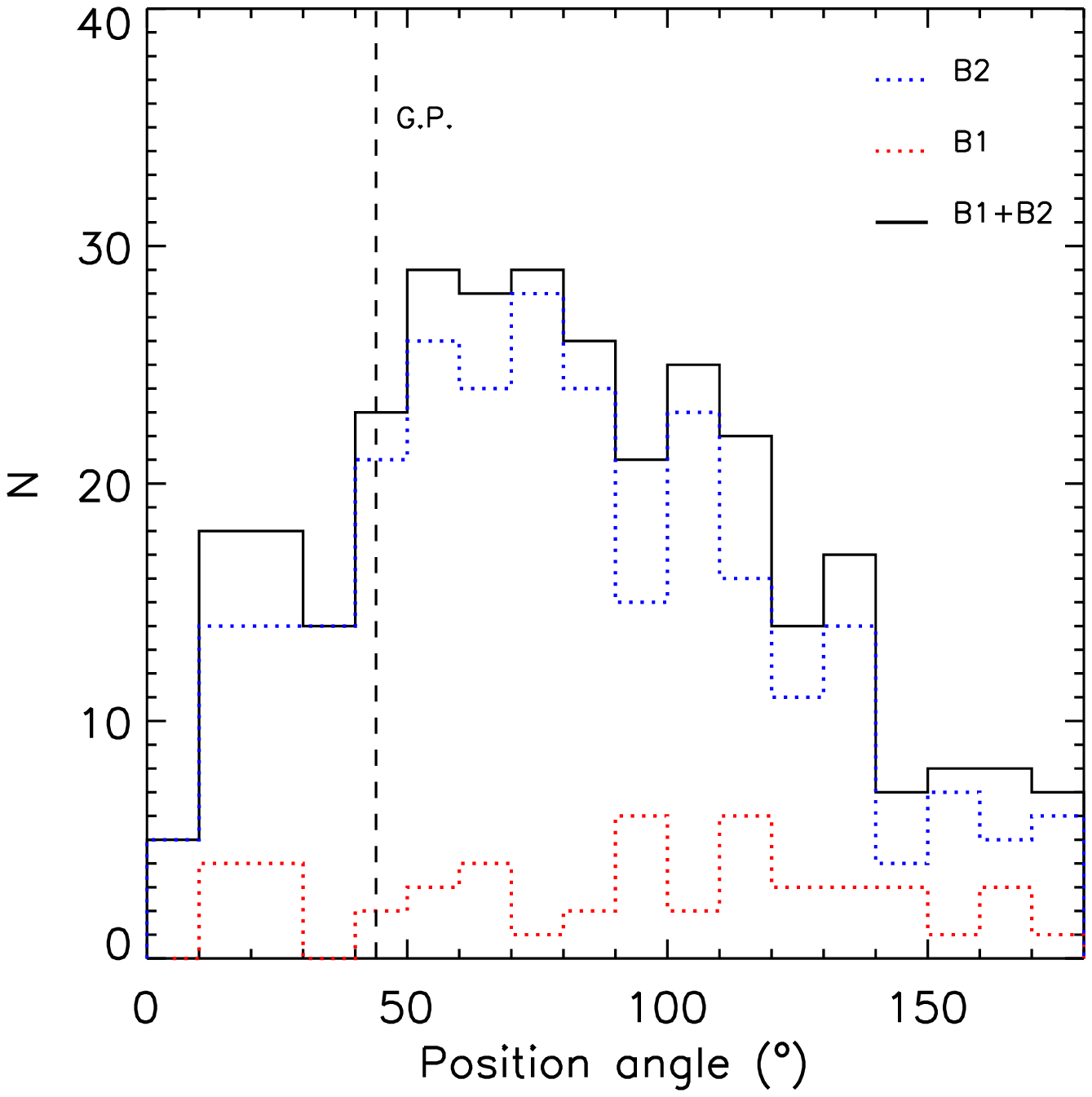}}
\caption{Distributions of position angles (after 90$^{\degree}$ rotation) corresponding to data with PI/$\delta PI >$ 2 shown as white vectors in figure \ref{Fig:pol_vec}.  Histograms are shown for Oph-B1 in red, Oph-B2 in blue, and the entire Oph-B clump in black. The bin size is 10$^{\degree}$ for each distribution. The dashed line represents the orientation of the Galactic Plane.}\label{Fig:comb_hist}
\end{figure}

Figure \ref{Fig:comb_hist} compares the distribution of position angle in Oph-B1 and Oph-B2 using the POL-2 data from figure \ref{Fig:pol_vec}. The figure shows histograms of each complex and of the entire Oph-B clump.

\subsection{Comparison of SCUBA-POL and SCUBA-2/POL-2 results}

\begin{figure*}%[t]
\includegraphics[width=90mm]{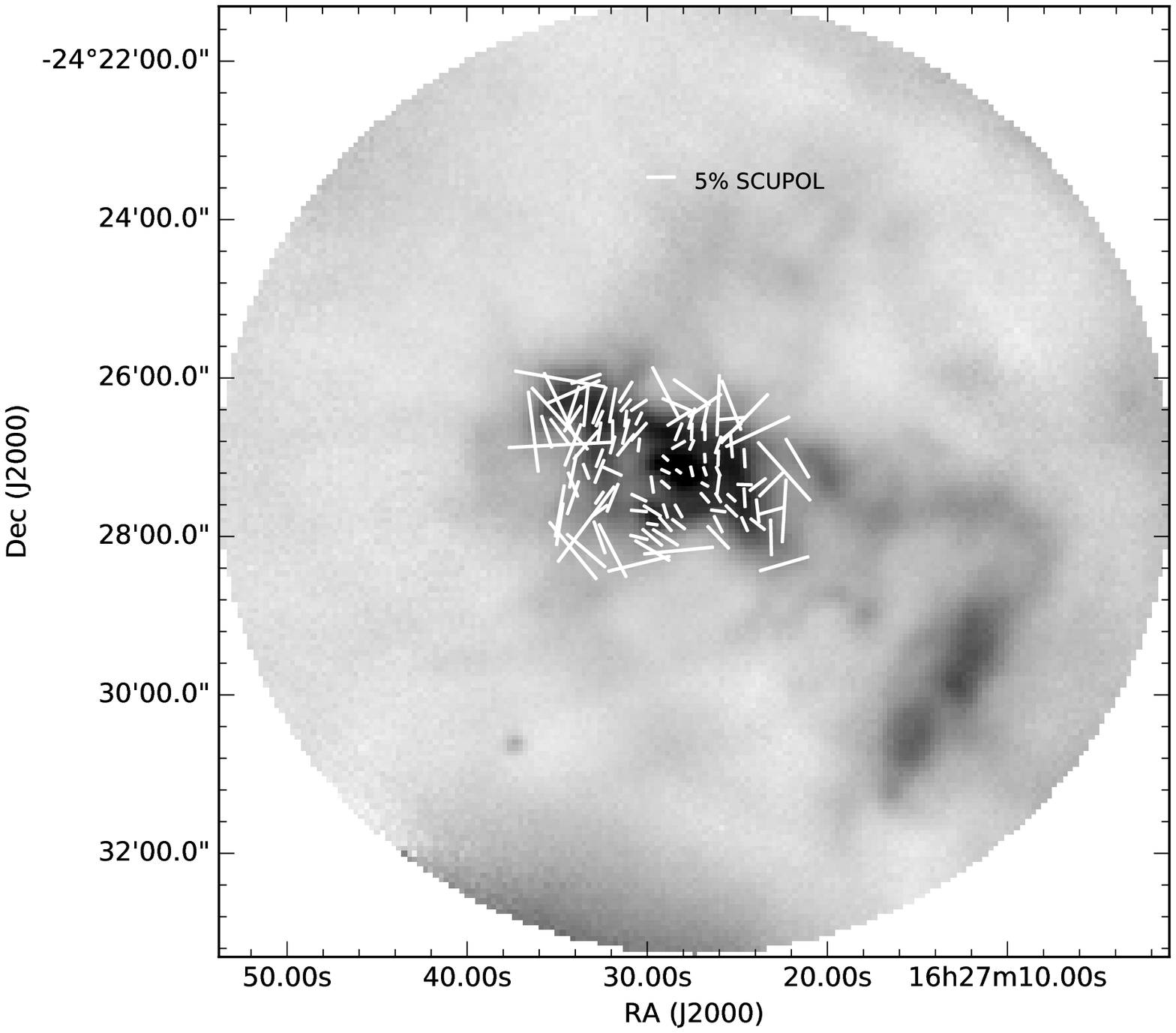}%
\includegraphics[width=90mm]{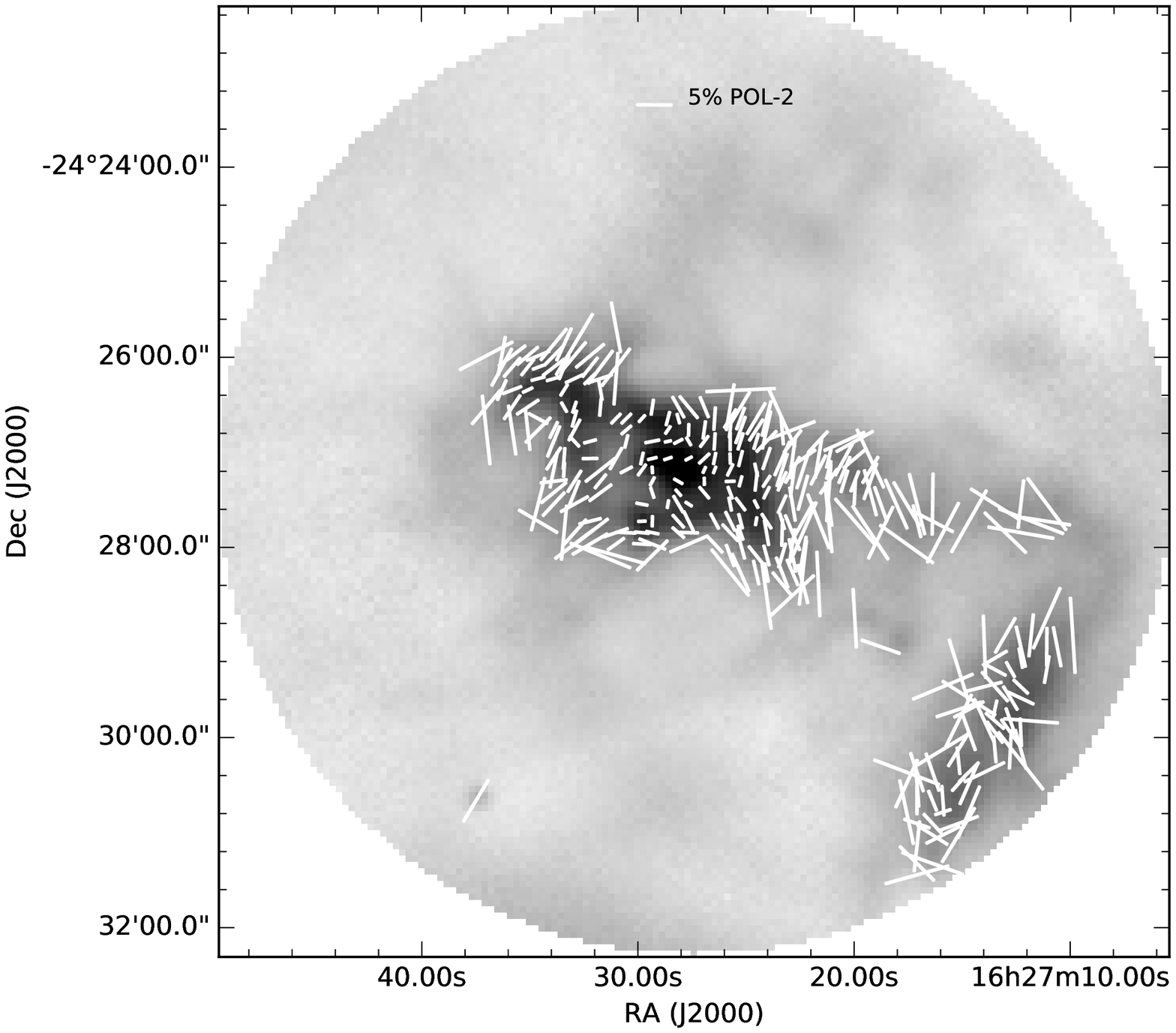}%
\caption{A comparison between SCUPOL (left panel) data \citep{2009ApJS..182..143M} and SCUBA-2/POL-2 (right panel) data (this work), both taken at the JCMT at 850$\mu$m. These are unrotated polarization vectors not B-fields orientations.}\label{Fig:pol_pol-2_comp1}
\end{figure*}

%----------------------------Horizontal layout--------------
%-----------separate---------------------------------------

\begin{figure}
\resizebox{8cm}{8cm}{\includegraphics{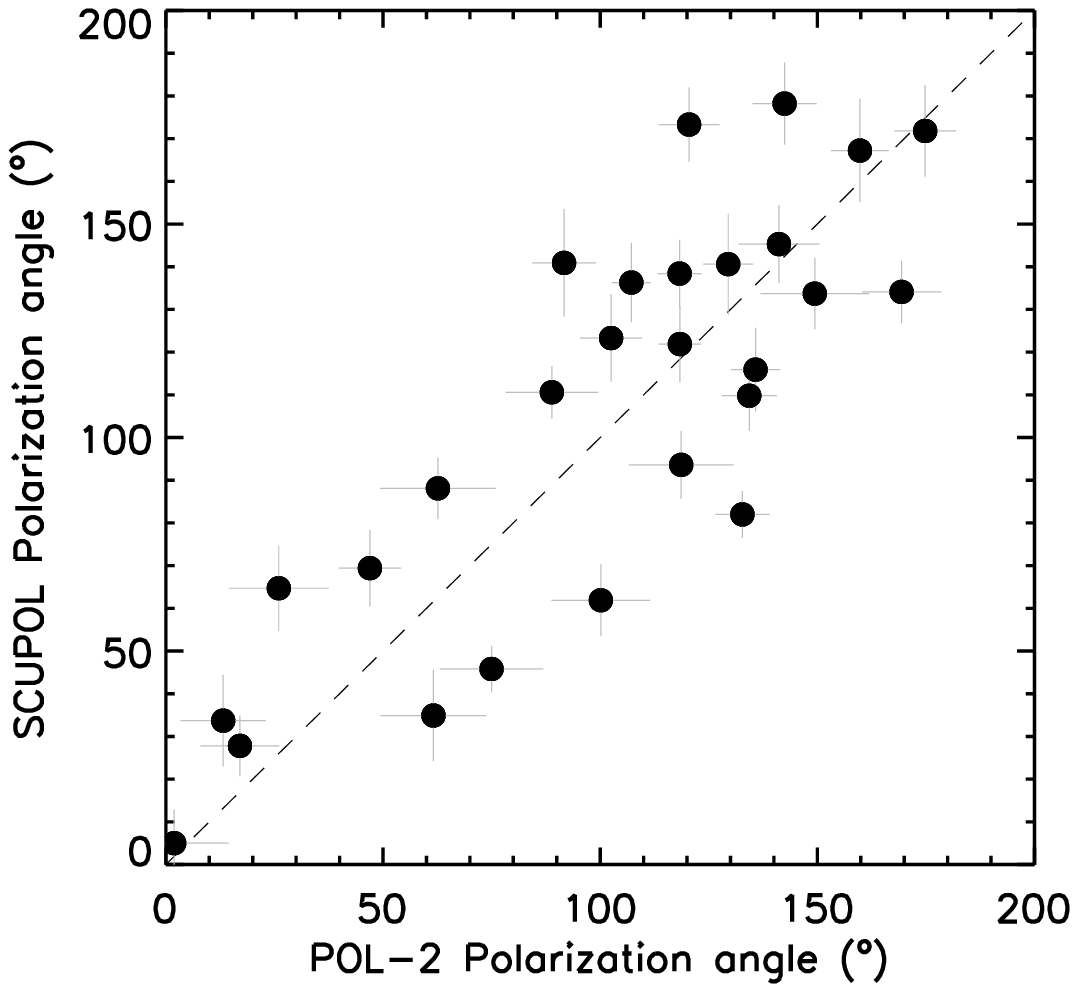}}\\
\resizebox{8cm}{8cm}{\includegraphics{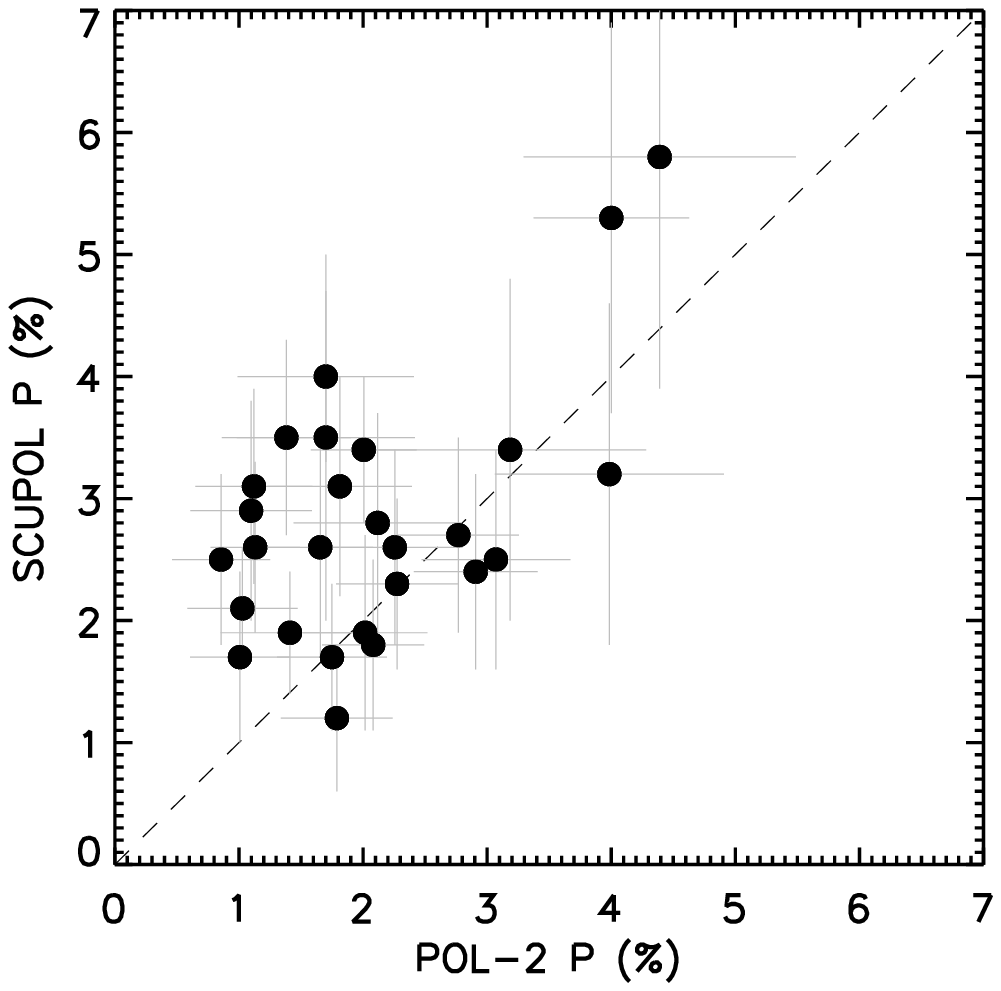}}
\caption{A comparison of polarization angle (top) and fraction (bottom) values obtained towards Oph-B2 by SCUPOL \citep{2009ApJS..182..143M} and SCUBA-2/POL-2 (this work). Dashed lines show a one-to-one relation (see text for details).}\label{Fig:pol_pol-2_comp2}
\end{figure}

\citet{2009ApJS..182..143M} cataloged polarization observations for various star-forming regions using 850$\mu$m data obtained using SCUBA-POL (SCUPOL) at the JCMT. Figure \ref{Fig:pol_pol-2_comp1} compares the polarization vectors (not rotated by 90$^{\degree}$) towards Oph-B2 (Oph-B1 was not observed with SCUPOL). We found 27 SCUPOL data points with coordinates which matched to those of our observations within 10$^{\prime\prime}$. The polarization vectors in common agree well in position angles. Figure \ref{Fig:pol_pol-2_comp2} shows a comparison of degree of polarization and position angle between the two datasets. The polarization angles measured by the two instruments agree with each other within the errors (figure \ref{Fig:pol_pol-2_comp2} upper panel). In contrast, the polarization fraction measured with SCUPOL are systematically higher than those with POL-2, especially in the low P values data (figure \ref{Fig:pol_pol-2_comp2} lower panel), suggesting that such a difference is caused by  the fact that errors in polarization fraction, P, do not obey Gaussian statistics, since P can never be negative, and hence, as discussed above, absolute values of P observed with all instruments and at all wavelengths tend to increase in areas of very low SNR. This has been discussed in detail by \citet{2006PASP..118.1340V} placing the confidence level in polarization measurements. The SCUPOL data have lower SNR than our own data, and consequently the polarization fractions measured with that instrument are typically higher than those measured with POL-2.

We also performed the Kolmogorov-Smirnov \citep[K-S; ][]{1992ApJ...385..416P} statistical test to quantify the deviation between SCUPOL and POL-2 polarization position angles towards Oph-B2 and found a probability of 0.905. The average difference in the orientation between the position angles of SCUPOL and POL-2 segments is found to be $\sim 20^{\degree}$. The distribution of the differences in the orientations of position angles in these two samples peaks at $\sim 10-20^{\degree}$. Detailed quantitative comparison of POL-2 data with SCUPOL is difficult because of the poor number statistics and low SNR in the SCUPOL data. However, we note that the two sets of vectors are broadly consistent, and agree best in the center of the SCUPOL field. Also, note that POL-2 data have much higher sensitivity than SCUPOL, which provide more reliable polarization angles.

%This difference is seen even in the case of high signal to noise ratio data with PI/$\delta$PI $>$ 3.

\section{Discussion} \label{sec:discuss}

\subsection{The magnetic field in NIR wavelengths}

\citet{2015ApJS..220...17K} presented the B-field morphology across Ophiuchus, measured using dust extinction in the J, H and K bands using SIRPOL\footnote{SIRIUS camera in POLarimeter mode on the IRSF 1.4 m telescope in South Africa \citep{2006SPIE.6269E..51K}}. Near-infrared (NIR) polarimetry can probe B-fields in diffuse environments ($\rm A_{V} \sim 10-20$ mag; \citealt{1999sf99.proc..212T, 2011ASPC..449..207T}). However, these data cannot probe the high-density regions where stars themselves form. To study the B-field morphology on $\lesssim$1 pc scales, \citet{2015ApJS..220...17K} compared the NIR data with optical polarimetry data tracing B-fields on 1-10 pc scales in the Ophiuchus region \citep{1976AJ.....81..958V}. Their investigation suggested that the B-field structures in the Ophiuchus clumps were distorted by the cluster formation in this region, which may have been induced by a shock compression by winds and/or radiation from the Scorpius$-$Centaurus association towards its west. The overall B-field observed at NIR wavelengths is found to have an orientation of 50$^{\degree}$ East of North \citep{2015ApJS..220...17K}.

\subsection{The magnetic field morphology in submm wavelengths}

Submm polarimetry probes B-fields in relatively dense ($\rm A_{V} \sim 50~mag$) regions of clouds. It can be noticed in figure \ref{Fig:comb_hist} that the distribution of position angles peaks at $\theta$= 50-80$^{\degree}$, suggesting that the large-scale B-field is a dominant component of the cloud's B-field geometry. There is a more clearly prevailing B-field orientation in Oph-B2, as demonstrated by the peak in the distribution shown in the histogram. The 1000-2000 AU-scale B-field traced by 850$\mu$m polarimetry seems to follow the structure of the two clumps, if we consider the field geometry as a whole. In figure \ref{Fig:pol_vec}, it is clear that a B-field component with an orientation of 50-80$^{\degree}$ is present in the low-density periphery of the cloud, and seems to be connecting well to the large-scale B-fields revealed at NIR wavelengths (shown as red vectors in the figure). This component is also present in Planck polarization observations \citep{2015A&A...576A.105P}. This orientation is also supported by the position angle of a single isolated detection in the south-east of the Oph-B region, which has a position angle $\sim $60$^{\degree}$. The mean magnetic field direction in Oph-B2 by Gaussian fitting to the position angles is measured to be 78$\pm$40$^{\degree}$. The position angle of Oph-B2 major axis is found to be 60$^{\degree}$ inferred by fitting an ellipse to the clump. The angular offset between the mean B-field direction and the major axis of Oph-B2 suggests that the B-field is consistent with a field that is parallel both to the long axis of Oph B2, and to the filament-like structure in which it lies. The diffuse filamentary structure connecting the two clumps seems to have B-field lines orthogonal to it. This can be seen from the overall geometry of the B-field lines in this low-density region, which follows the clump structures. We more clearly show the smoothed B-field geometry in figure \ref{Fig:binned}. In this figure, the B-field vectors corresponding to P$>$5\% and P$<$5\% are plotted in blue and white, respectively. The overall B-field geometry seems somewhat disordered (see section 4.4), with the blue vectors showing a higher degree of polarization towards the more diffuse parts of the cloud, and the white vectors showing the B-fields in the dense core regions.

%*****************************************************************************************************************

\subsection{Variation of polarization fraction}

\begin{figure*}%[t]
\includegraphics[width=90mm]{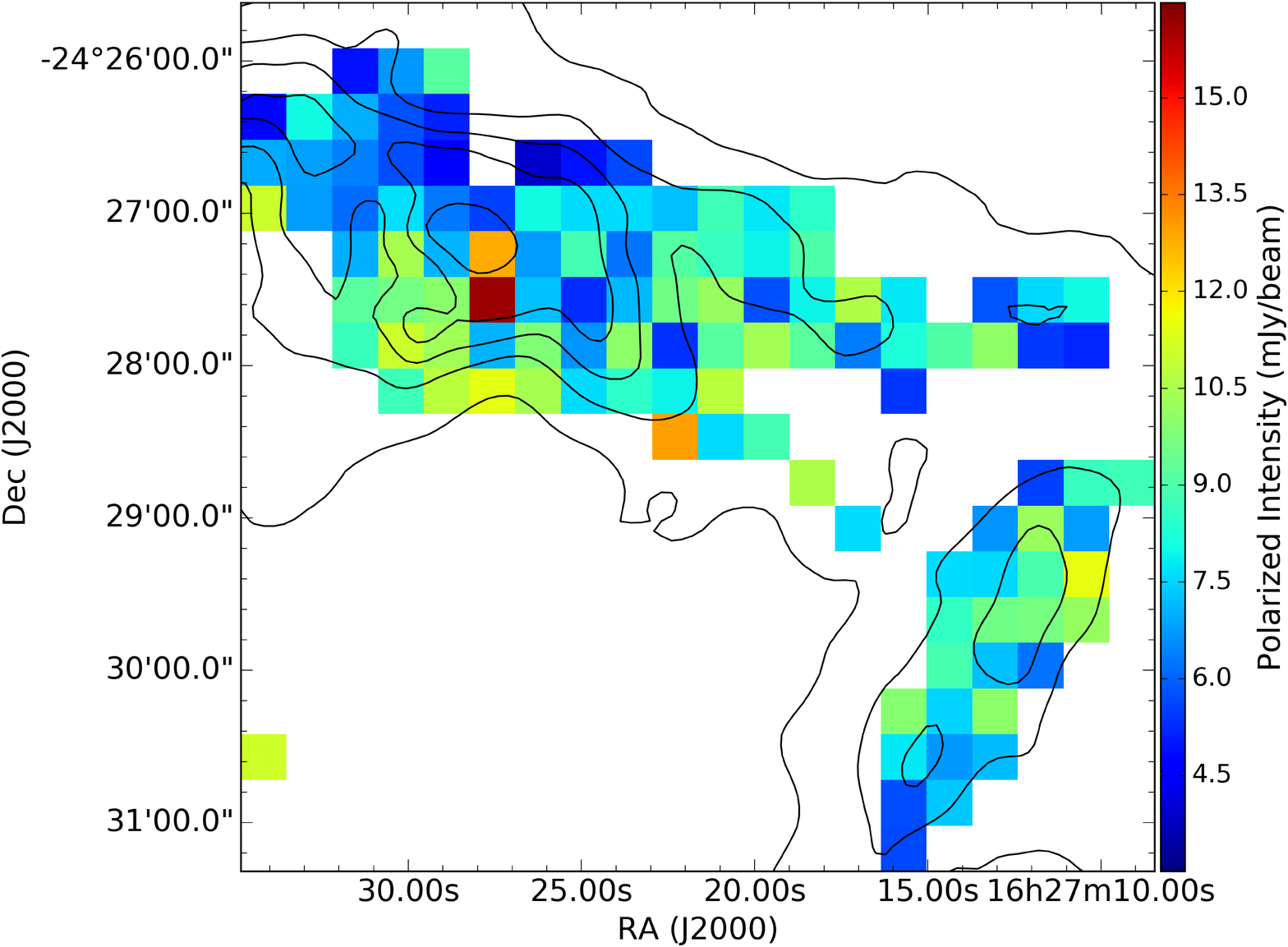}%
\includegraphics[width=90mm]{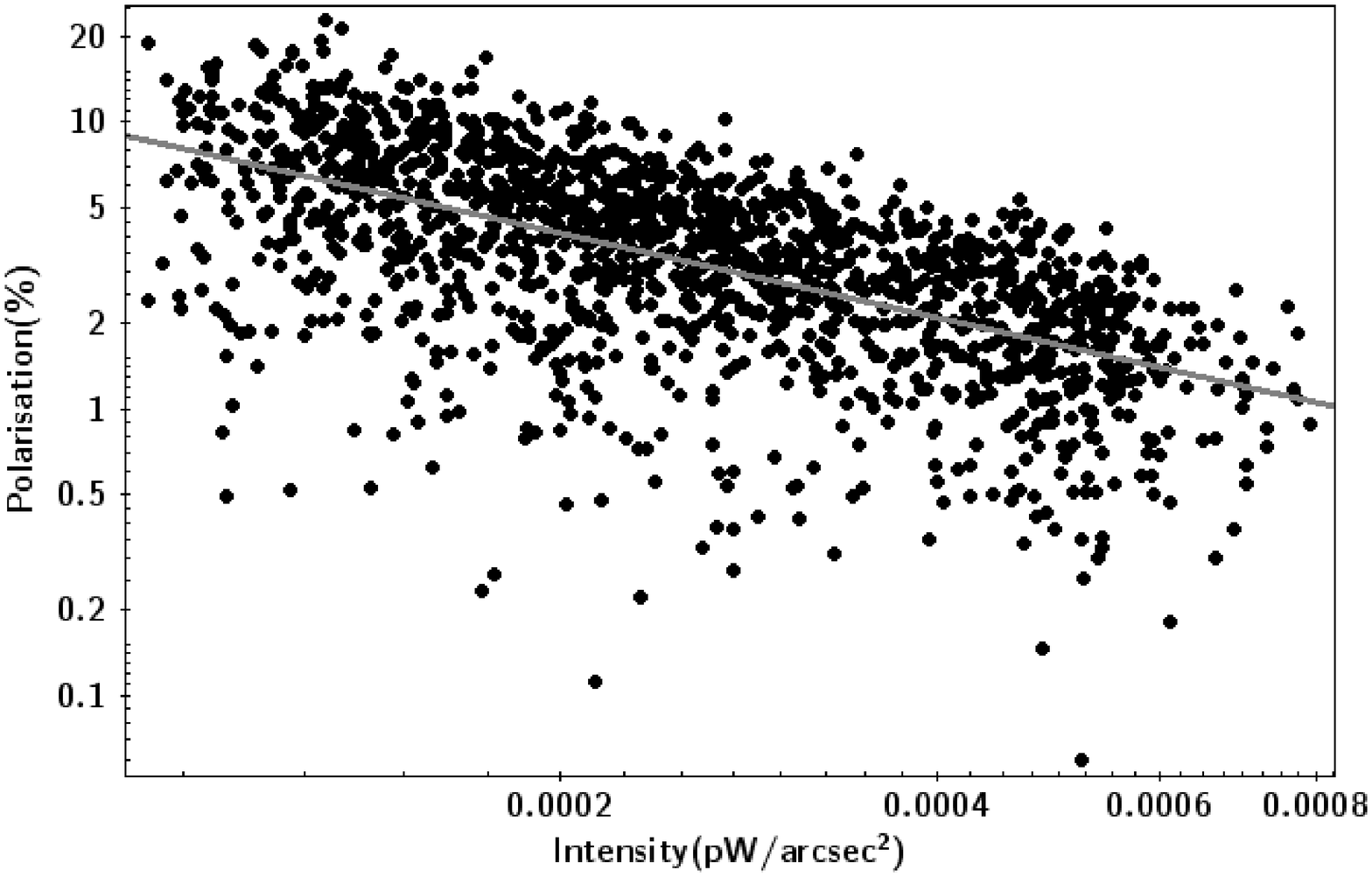}%
\caption{Left panel shows the map of polarization intensity with dust continuum contour overlaid. The degree of polarization vs. total intesity plot is shown in right panel after selecting the sample using I $>$ 0, $\frac{I}{\delta I}$ $>$ 20, and P $>$ 0. The grey line shows the linear fitting in the data.}\label{Fig:p_vs_int}
\end{figure*}

The left-hand panel in figure \ref{Fig:p_vs_int} shows a map of polarization intensity towards Oph-B, with contours of Stokes I emission overlaid. The figure shows a decrease in polarization fraction in denser regions (as traced by the Stokes I emission).  This depolarization has also been seen in many previous studies \citep[e.g.][]{1996ApJ...470..566D, 2002ApJ...574..822M, 2006Sci...313..812G, 2009ApJ...695.1399T, 2009ApJ...700..251T, 2013ApJ...763..135T}. It is noticed that the polarization fraction is relatively higher on the edges of the cloud than in the denser core regions (see figure \ref{Fig:binned}).  Plausible causes for the polarization fraction hole effect have been discussed in previous studies, such as \citet{2014ApJS..213...13H}.  They found that in low-resolution maps (i.e $\sim$20$^{\prime\prime}$ resolution, obtained from SHARP\footnote{The SHARC-II polarimeter for CSO}, Hertz and SCUBA), depolarization resulted from unresolved structure being averaged across the beam. However, in high-resolution maps (i.e. $\sim$2.5$^{\prime\prime}$ resolution, obtained using CARMA\footnote{Combined Array for Research in Millimeter Astronomy}), depolarization persisted even after resolving the twisted plane-of-sky B-field morphologies.

Except for a very few lines of sight through the densest parts of protostellar disks, (sub)millimeter-wavelength thermal dust emission is optically thin. Therefore, polarization data at these wavelengths will be an integration of the emission along the entire line of sight. If the B-field changes along the line of sight (e.g. due to turbulence), the observed dust polarization will be averaged, and so will result in a lower observed polarization fraction. Future higher-resolution ALMA observations can help in tracing the magnetic field structure at different scales in the cloud \citep[e.g.,][]{2013ApJ...768..159H, 2014ApJS..213...13H}. 

The polarization hole effect has also been explained as resulting from different grain characteristics in dense cores than in diffuse regions. Grain growth can make grains more spherical and thus harder to align with B-fields by radiative torques. Grain alignment can also be disturbed due to collisions in the higher density regions resulting in low polarization.

Alternatively the polarization hole effect could arise from a loss of photon flux to drive grain alignment in very dense regions. According to modern grain alignment theory, dust grains are aligned by radiative torques \citep{2007MNRAS.378..910L, 2008MNRAS.388..117H, 2009ApJ...697.1316H, 2016ApJ...831..159H}. Towards a dense region, the magnitude of these Radiative Alignment Torques (RATs) decreases due to the attenuation of the interstellar radiation. As a result, grain alignment may be suppressed in very high extinction regions, resulting in a decrease in the observed fractional polarization.

\citet{2003ApJ...592..233W} found that the polarization fraction in molecular clouds follows the relation P$\rm \propto I^{-\alpha}$, with $\alpha \sim 0.5 -1.2$. They also reported a value of $\alpha$ of -0.43 towards some Bok globules. \citet{1998ApJ...499L..93A} observed a similar behavior in the dust extinction polarization of background starlight. \citet{2008ApJ...679..537F} performed a statistical investigation of the previous studies in order to understand this trend. They produced synthetic polarimetric maps and found an anti-correlation between polarization degree and column density with a value of $\alpha$ of -0.5 (assuming perfect grain alignment with no dependence on extinction), due to an increase in the dispersion of the polarization angle along the line of sight in the denser regions. \citet{2005ApJ...631..361C}, in their study of grain alignment, claimed that power law index is sensitive to the grain size distribution.

Measured values of $\alpha$ can vary in different cases. For example, \citet{2000ApJ...531..868M} found a relation of P$\rm \propto I^{-0.7}$ towards OMC-3, but P$\rm \propto I^{-0.8}$ in the Barnard 1 cloud. In the case of dense star-forming cores, \citet{2001ApJ...561..871H} measured P$\rm \propto I^{-0.6}$, whereas \citet{2002ApJ...566..925L} found P$\rm \propto I^{-0.8}$, and \citet{2004Ap&SS.292..225C} obtained a slightly steeper value of $\alpha$, with P$\rm \propto I^{-1.2}$. The relationship between polarization fraction and intensity may be different in dense cores. \citet{2017ApJ...847...92H} found a relatively high polarization fraction near the total intensity peak of Serpens SMM1 using ALMA data. This higher degree of polarization may possibly be due to additional radiative torques from its young central protostar. 

The right-hand panel of figure \ref{Fig:p_vs_int} compares degree of polarization with 850$\mu$m dust emission intensity in Oph-B. We measured the power-law slope in the distribution using a least-squares fit, and found $\alpha \approx -0.9$. A similar value has been reported by \citet{2014A&A...569L...1A} towards starless cores. Most recent ALMA observations at 0.26$^{\arcsec}$ resolution towards W51 e2, e8, North show a slope of -0.84 to -1.02 and SMA observations at 2$^{\arcsec}$ resolution towards the same source in e2/e8 and North reported a slope of -0.9 and -0.86, respectively \citep{2018ApJ...855...39K}. Higher-resolution polarization observations (i.e. using ALMA) are needed to probe the magnetic fields toward the very centers of the embedded cores in Oph-B, on the scales $<$ 1000 AU.

%***************************Written by Kate********************************************

\subsection{The Davis-Chandrasekhar-Fermi analysis for B-field strength}

We use the Davis-Chandresekhar-Fermi (DCF; \citealt{PhysRev.81.890.2D, 1953ApJ...118..113C}) method to estimate the plane-of-sky magnetic field strength in Oph-B.  This method assumes that the underlying magnetic field geometry of the region under consideration is uniform, and hence that the observed dispersion in position angle is a measure of the distortion in the field geometry caused by turbulence, and also that the distribution of vectors about the mean field direction is approximately Gaussian, and thus well-characterized by its standard deviation. 

The DCF method estimates the plane-of-sky B-field ($B_{pos}$) using the formula

\begin{equation}
B_{pos} = Q\sqrt{4\pi \rho}\frac{\sigma_{v}}{\sigma_{\theta}} , 
\end{equation}
where $\rho$ is the gas density, $\sigma_{v}$ is the 1D non-thermal velocity dispersion of the gas, $\sigma_{\theta}$ is the dispersion in polarization angle, and $Q$ is a factor of order unity that accounts for variations in the B-field on scales smaller than the beam.

\citet{2004Ap&SS.292..225C} formulated this as:

\begin{equation}
B_{pos} \approx 9.3\sqrt{n({\rm H}_{2})}\frac{\Delta v}{\sigma_{\theta}}\,\mu{\rm G},
\end{equation}
where $n({\rm H}_{2})$ is the number density of molecular hydrogen in cm$^{-3}$, $\Delta v$  = $\sigma_{v}\sqrt{8\ln 2}$ in km\,s$^{-1}$, $\sigma_{\theta}$ is in degrees, and $Q$ is taken to be 0.5 (cf. \citealt{2001ApJ...546..980O}).

\subsubsection{Number density}

\citet{1998A&A...336..150M} infer a volume density in Oph-B of $n({\rm H}_{2}) = 3.2\times 10^{6}$\,cm$^{-3}$ from 1.3\,mm dust continuum measurements.  We adopt this value throughout the subsequent analysis.  These authors estimate that their measured column densities are accurate to within a factor $\lesssim 2$, so we conservatively adopt a fractional uncertainty of $\pm 50$\,\% on their stated volume density.

\subsubsection{Velocity dispersion}

\citet{2007A&A...472..519A} observed Oph-B in the N$_{2}$H$^{+}$ $J=1\to 0$ line with beam size 26${\arcsec}$ using IRAM 30m telescope.  \citet{2015MNRAS.450.1094P} showed that there is good agreement between core masses derived using these N$_{2}$H$^{+}$ data and core masses derived using SCUBA-2 850$\rm \mu$m continuum observations of Oph-B (see their figure 8). Hence, we assume that N$_{2}$H$^{+}$ traces the same density region as our SCUBA-2/POL-2 observations do, and so that the non-thermal velocity dispersions measured by \cite{2007A&A...472..519A} are representative of the non-thermal motions perturbing the magnetic field which we observe in Oph-B.

\citet{2007A&A...472..519A} list non-thermal velocity dispersion for 19 of the cores in Oph-B2 identified by \citet{1998A&A...336..150M} in their Table 4 (note that we use the non-background-subtracted values in order to have the best possible correspondence with the integrated dust column).  Averaging these velocity dispersion, we estimate $\sigma_{v} = 0.24\pm 0.08$\,km\,s$^{-1}$, i.e. $\Delta v = 0.56\pm 0.20$\,km\,s$^{-1}$, where the uncertainties give the standard deviation on the mean.

\subsubsection{Angular dispersion}

\begin{figure*}
  \centering
  \includegraphics[width=\textwidth]{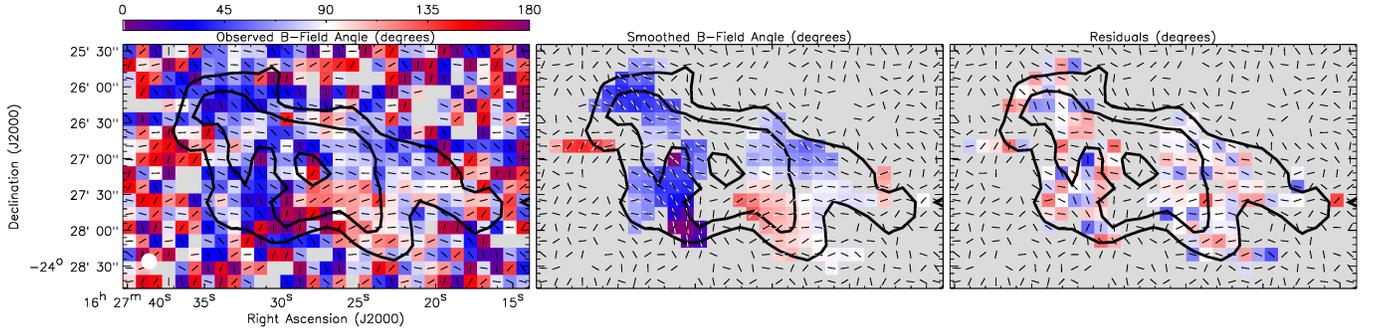}
  \caption{Left panel: The observed magnetic field angles in Oph-B2. Middle panel: The magnetic field angle map smoothed with a 3$\times$3-pixel boxcar filter.  Right panel: The residual map from subtracting the smoothed map from the observations. Contours show Stokes I emission at 1.0, 2.0 and 3.0 mJy/$\rm arcsec^{2}$}\label{Fig:boxcar}
\end{figure*}

To ensure that we are measuring the dispersion of statistically independent pixels, and to maximize the fraction of the map with high SNR measurements, we gridded the vector map to 12\arcsec (approximately beam-sized) pixels. We performed the DCF analysis on Oph-B2 only, because the field in Oph-B1 is quite disordered, as seen in figure \ref{Fig:comb_hist}, with no clear peak in the distribution of angles. The DCF method thus cannot be meaningfully applied in this region.

There are many methods of estimating the angular dispersion in polarization maps. The polarization vectors in Oph-B2 show a relatively ordered morphology with some scatter along the edges. Naively taking the standard deviation of all pixels in Oph-B2 with $PI/\delta PI > 2$, we measure $\sigma_{\theta} = 42^{\circ}$. This value is significantly larger than the maximum value at which the standard DCF method can be safely applied ($\lesssim 25^{\circ}$; \citealt{2001ApJ...561..800H}). This large standard deviation includes both the scatter in polarization vectors and any underlying variations in the ordered field.

Instead, we followed the technique to measure the dispersion of polarization angle presented by \citet{2017ApJ...846..122P}.  They used an unsharp-masking method to remove the underlying ordered magnetic field structure from their observations.  Firstly, they estimated the general magnetic field structure by smoothing their map of magnetic field with a $3\times 3$-pixel boxcar filter. Secondly, they subtracted the smoothed map from the observed magnetic field angle map.  The resulting residual map was taken to represent the non-ordered component of the magnetic field.  Finally, they used the standard deviation in the residual map to represent the standard deviation in the magnetic field angle, $\sigma_{\theta}$. We refer readers to \citet{2017ApJ...846..122P} for a detailed description of the methodology.

\begin{figure}
 \centering
  \includegraphics[width=0.5\textwidth]{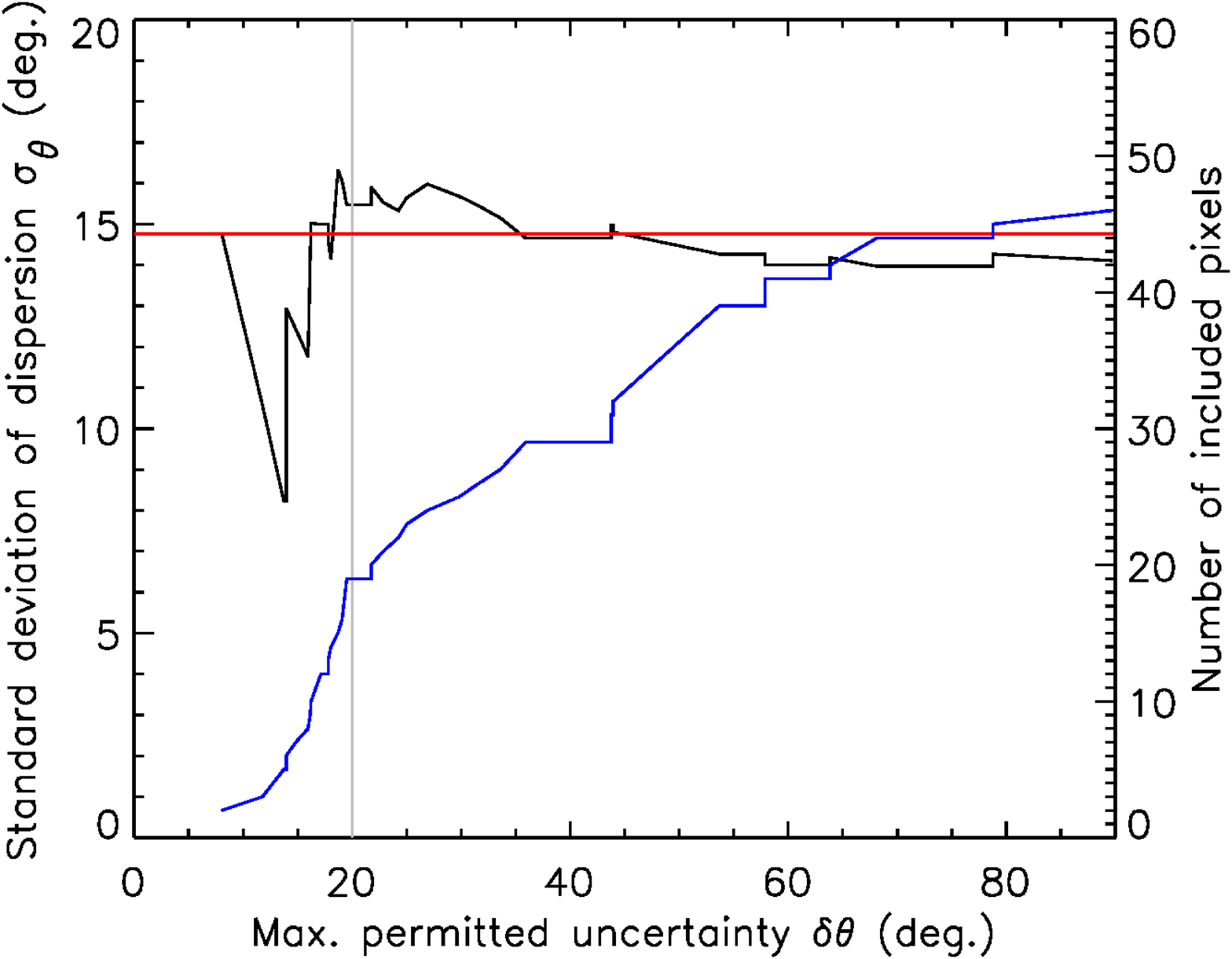}
  \caption{Black line: the standard deviation ($\sigma_{\theta}$), of the distribution of the magnetic field angles in Oph-B2 about the mean magnetic field direction as a cumulative function of maximum uncertainty in the $3\times 3$-pixel smoothing box (left-hand axis).  Red line: the mean value of standard deviation, measured over the distribution to the right of the grey line (left-hand axis).  Blue line: the number of pixels included in the cumulative distribution (right-hand axis).}\label{Fig:cumulat_OphB}
\end{figure}

We applied the \citet{2017ApJ...846..122P} method to our Oph-B2 data.  Figure \ref{Fig:boxcar} shows the observed projected magnetic field (left panel), the inferred ordered component to the field (middle panel), and the residual map (right panel).  We restrict our analysis to pixels where $P/\delta P>3$ (although the SNR may be lower in the surrounding pixels, which contribute to the mean field estimate), and exclude any pixel where the maximum difference in angle between pixels within the boxcar filter is $\geq 90$ degrees, on the assumption that the field in the vicinity of that pixel is not sufficiently uniform for the smoothing function to be valid.  We measure a standard deviation in the residual map of $\sigma_{\theta} = 14.8^{\circ}\pm 0.7^{\circ}$, as shown in figure \ref{Fig:cumulat_OphB}.

The method proposed by \citet{2017ApJ...846..122P} was developed for data with higher SNR than we achieve in Oph-B.  We tested the accuracy of this estimate of $\sigma_{\theta}$ by measuring the standard deviation in angle in small regions of Oph-B2 which do not show significant ordered variation across them.  We measured $\sigma_{\theta} = 12.8^{\circ}$ in the north-west of Oph-B2, $\sigma_{\theta} = 16.6^{\circ}$ in the north-east, and $\sigma_{\theta} = 16.5^{\circ}$ in the south-west.  This suggests that the dispersion in angle that we measure using the unsharp masking method is representative of the true dispersion in angle may be due to non-thermal motions in Oph-B2, and so we adopt $\sigma_{\theta}\sim 15^{\circ}$ going forward.

\subsubsection{Magnetic field strength}

Using our values of $n({\rm H}_{2}) = (3.2\pm 1.6)\times 10^{6}$\,cm$^{-3}$, $\Delta v = 0.56\pm 0.20$\,km\,s$^{-1}$ and $\sigma_{\theta}= 14.8\pm0.7^{\circ}$, we estimate a plane-of-sky magnetic field strength in Oph-B2 of 630\,$\rm \mu$G.  Combining our uncertainties using the relation
\begin{equation}
 \frac{\delta B_{pos}}{B_{pos}} = \frac{1}{2}\frac{\delta n({\rm H}_{2})}{n({\rm H}_{2})} + \frac{\delta \Delta v}{\Delta v} + \frac{\delta\sigma_{\theta}}{\sigma_{\theta}} ,
\end{equation}

where $\delta n({\rm H}_{2})$ represents the uncertainty on $n({\rm H}_{2})$ and so forth, we find a total fractional uncertainty of $\delta B_{pos}/B_{pos}\approx 66$\,\%.  We note that a significant fraction of the contribution to this uncertainty is systematic, and thus that our result of $\rm B_{sky} = 630 \pm 410\,\mu$G suggests that the magnetic field strength in Oph-B2 is possibly within the range $\rm 200-1000\mu$G.

\subsubsection{Small-angle approximation}

The DCF method fails at large angular dispersions in part due to the failure of the small-angle approximation, as discussed in detail by \citet{2001ApJ...561..800H} and \citet{2008ApJ...679..537F}.  For our measured value of angular dispersion, $\sigma_{\theta} = 15^{\circ} = 0.262$ radians, $\tan(\sigma_{\theta}) = 0.268$, i.e. $\tan(\sigma_{\theta})/\sigma_{\theta}=1.02$.  We are in the regime in which the small-angle approximation holds well, and so the standard DCF method is an appropriate choice here.

\subsubsection{Line-of-sight turbulence}

Many authors have discussed the appropriate value of the $Q$ parameter in the DCF equation to properly account for variations if the magnetic field along the line of sight or field on scales smaller than the beam (e.g. \citealt{2001ApJ...546..980O}; \citealt{2001ApJ...561..800H}; \citealt{2009ApJ...706.1504H, 2016ApJ...821...21C}). Modeling suggests that the DCF method will overestimate the magnetic field strength by a factor of $\sqrt{N}$, where $N$ is the number of turbulent eddies along the line of sight \citep{2001ApJ...561..800H}.  \citet{2016ApJ...821...21C} propose that the number of turbulent eddies can be estimated using the relation
\begin{equation}
  \frac{1}{\sqrt{N}} \sim \frac{\sigma_{V_{c}}}{\sigma_{v}},
\end{equation}
where $\sigma_{V_{c}}$ is the standard deviation in the centroid velocities across the observed region, and $\sigma_{v}$ is the average line-of-sight non-thermal velocity dispersion, as previously discussed.

Using the centroid velocities listed in Table 2 of \cite{2007A&A...472..519A}, we estimate $\sigma_{V_{c}} = 0.218$\,km\,s$^{-1}$, for the same 19 cores as were used to estimate $\sigma_{v}$.  Thus, we find that $\sigma_{V_{c}}/\sigma_{v} = 0.915$, i.e. $N\sim 1$. Thus we do not expect large numbers of turbulent eddies along the line of sight, and so any over-estimation of the magnetic field in Oph-B2 due to this effect should be minimal.

%******************************************************************************************************

\subsubsection{Mass-to-flux ratio}
The relative importance of gravity and B-fields to the overall stability of a region is represented by the mass-to-flux ratio, $\lambda$ \citep{2004Ap&SS.292..225C}. The stability of cores can be estimated observationally using $\rm H_{2}$ column density and B-field strength. We assume that the average field strength from the larger Oph-B2 region applies to the dense cores within the region. \citet{1998A&A...336..150M} estimated the column density of Oph-B2 to be $\mathrm{41\times 10^{22} cm^{-2}}$. Thus, its mass-to-flux stability parameter, $\lambda$, can be estimated using the relation given by \citet{2004Ap&SS.292..225C},

\begin{equation}\label{lambda}
\mathrm{\lambda = 7.6 \times 10^{-21} \frac{N(H_{2})}{B_{pos}}} ,
\end{equation}
 where $\rm N(H_{2})$ is the molecular hydrogen column density in $\mathrm{cm^{-2}}$ and $\mathrm{B_{pos}}$ is the plane-of-the sky field strength in $\mu$G. From our values of $\mathrm{B_{pos}}$ (see Section 4.4) and $\rm N(H_{2})$, we find $\lambda=$4.9$\pm$3.2 where the error comes from the uncertainty in B-field strength measurement only. According to \citet{2004Ap&SS.292..225C} this value can be overestimated by a factor of 3 due to geometric effects. Thus, the corrected value of $\lambda$ is 1.6$\pm$1.1, which is treated as the lower limit of $\lambda$ in this work. A value of $\lambda >$ 1 indicates that the core is magnetically supercritical, suggesting that its B-field is not sufficiently strong to prevent gravitational collapse. The values of $\lambda$ estimated towards Oph-B2 suggest that this clump is slightly magnetically supercritical.

\subsection{Relative orientation between outflows and magnetic fields}

\citet{2014ApJS..213...13H} used CARMA observations to investigate the relationship between outflow directions and B-field morphologies in a large sample of young star-forming systems.  They found that their sample was consistent with random alignments between B-fields and outflows.  Nevertheless, sources with low polarization fractions (similar to Oph-B2) appear to have a slight preference for misalignment between B-field directions and outflow axes, suggesting that their field lines are wrapping around the outflow due to envelope rotation.

Oph-B2 contains a prominent outflow driven by IRS47 \citep{2015MNRAS.447.1996W}. The plane-of-sky position angle of outflow is $\sim$140$^{\degree}$, which is roughly 50$^{\degree}$ offset from the general orientation of the B-field in the vicinity of the outflow. The offset between the overall B-field orientation in Oph-B2 ($\sim$80$^{\degree}$) and the outflow from IRS47 is $\sim$60$^{\degree}$. These results suggest that outflows are not aligned with the B-field in Oph-B on the size scales that we observe. This is consistent with models which suggest outflows are misaligned to magnetic field orientation \citep{2009A&A...506L..29H, 2012A&A...543A.128J, 2013ApJ...774...82L}.

\section{Summary} \label{sec:conclusion}
This paper presents the first-look results of SCUBA-2/POL-2 observations in the 850$\mu$m waveband towards the Oph-B region by the BISTRO survey. In this study, we mainly focused on the role of B-fields in star formation. The main findings of this work are summarized below:

\begin{enumerate}

\item We found polarization fractions and angles to be consistent with previous SCUPOL data. But our POL-2 map is more sensitive than the archival data and covers a larger area of the Oph-B clump.

\item We identify distinct magnetic field morphologies for the two sub-clumps in the region, Oph-B1 and Oph-B2.  Oph-B1 appears to be consistent with having a random magnetic field morphology on the scales which we observe, whereas Oph-B2 shows an underlying ordered field in its denser interior.

\item The field in Oph-B2 is relatively well-ordered and lies roughly parallel ($\sim$18$^{\degree}$) to the long axis of the clump. 

\item We find that the B-field component lying on the low density periphery of the Oph-B2 clump is parallel to the large-scale B-field orientation identified in previous studies using observations made at NIR wavelengths. The sub-mm data show detailed structure that could not be detected with the NIR data.

\item The decrease in the polarization fraction is observed in the densest regions of both Oph-B1 and B2.

\item  We measured the magnetic field strength in Oph-B2 using the Davis-Chandrasekhar-Fermi (DCF) method.  Since this sub-clump has an underlying ordered field structure, we used an unsharp-masking technique to obtain an angular dispersion of $\sim15^{\degree}$.  We measured a field strength of 630$\pm$410$\mu$G and a mass-to-flux ratio of 1.6$\pm$1.1, suggesting that Oph-B2 is slightly magnetically supercritical.
  
\item We examined the relative orientation of the most prominent outflow in Oph-B2 (IRS 47) and the local magnetic field.  The two have orientations that are offset by $\sim$50$^{\degree}$, suggesting they are not aligned. The angular offset between the large-scale magnetic field in Oph-B2 and the outflow orientation is $\sim$60$^{\degree}$, suggesting consistency with models which predict that outflows should be misaligned to magnetic field direction.

\end{enumerate}

\section{Acknowledgments}
Authors thank referee for a constructive report that helped in improving the content of this paper. The James Clerk Maxwell Telescope is operated by the East Asian Observatory on behalf of the National Astronomical Observatory of Japan, the Academia Sinica Institute of Astronomy and Astrophysics, the Korea Astronomy and Space Science Institute, the National Astronomical Observatories of China and the Chinese Academy of Sciences (Grant No. XDB09000000), with additional funding support from the Science and Technology Facilities Council of the United Kingdom and participating universities in the United Kingdom and Canada. The James Clerk Maxwell Telescope has historically been operated by the Joint Astronomy Centre on behalf of the Science and Technology Facilities Council of the United Kingdom, the National Research Council of Canada and the Netherlands Organisation for Scientific Research. Additional funds for the construction of SCUBA-2 and POL-2 were provided by the Canada Foundation for Innovation. The data used in this paper were taken under project code M16AL004. AS would like to acknowledge the support from KASI for postdoctoral fellowship. K.P. acknowledges support from the Science and Technology Facilities Council (STFC) under grant number ST/M000877/1 and the Ministry of Science and Technology, Taiwan, under grant number 106-2119-M-007-021-MY3, and was an International Research Fellow of the Japan Society for the Promotion of Science for part of the duration of this project. CWL and MK were supported by Basic Science Research Program through the National Research Foundation of Korea (NRF) funded by the Ministry of Education, Science and Technology (CWL: NRF-2016R1A2B4012593) and the Ministry of Science, ICT \& Future Planning (MK: NRF-2015R1C1A1A01052160). W.K. was supported by Basic Science Research Program through the National Research Foundation of Korea (NRF-2016R1C1B2013642). JEL is supported by the Basic Science Research Program through the National Research Foundation of Korea (grant No. NRF-2018R1A2B6003423) and the Korea Astronomy and Space Science Institute under the R\&D program supervised by the Ministry of Science, ICT and Future Planning. D.L. is supported by NSFC No. 11725313. S.P.L. acknowledges support from the Ministry of Science and Technology of Taiwan with Grant MOST 106-2119-M-007-021-MY3. K.Q. acknowledges the support from National Natural Science Foundation of China (NSFC) through grants NSFC 11473011 and NSFC 11590781. This research has made use of the NASA Astrophysics Data System. AS thanks Maheswar G. for discussion on polarization in star-forming regions. AS also thanks Piyush Bhardwaj (GIST, South Korea) for a critical reading of the paper. The authors wish to recognize and acknowledge the very significant cultural role that the summit of Maunakea has always had within the indigenous Hawaiian community, with reverence. AS specially thanks the JCMT TSS for the hard-earned data.

%======================================================================================
\bibliographystyle{aasjournal}
\bibliography{references}

\end{document}